# Metastability of quantum droplet clusters

Yaroslav V. Kartashov,[1,2] Boris A. Malomed,[3] and Lluis Torner[1,4]

[1]*ICFO-Institut de Ciencies Fotoniques, The Barcelona Institute of Science and Technology, 08860 Castelldefels (Barcelona), Spain*
[2]*Institute of Spectroscopy, Russian Academy of Sciences, Troitsk, Moscow, 108840, Russia*
[3]*Department of Physical Electronics, School of Electrical Engineering, Faculty of Engineering, and Centre for Light-Matter Interaction, Tel Aviv University, 69978 Tel Aviv, Israel*
[4]*Universitat Politecnica de Catalunya, 08034, Barcelona, Spain*

We show that metastable ring-shaped clusters can be constructed from two-dimensional quantum droplets in systems described by the Gross-Pitaevskii equations augmented with Lee-Huang-Yang quantum corrections. The clusters exhibit dynamical behaviours ranging from contraction to rotation with simultaneous periodic pulsations, or expansion, depending on the initial radius of the necklace pattern and phase shift between adjacent quantum droplets. We show that, using an energy-minimization analysis, one can predict equilibrium values of the cluster radius that correspond to rotation without radial pulsations. In such a regime, the clusters evolve as metastable states, withstanding abrupt variations in the underlying scattering lengths and keeping their azimuthal symmetry in the course of evolution, even in the presence of considerable perturbations.

PhySH Subject Headings: Solitons; Superfluids; Mixtures of atomic and/or molecular quantum gases

Among the variety of possible self-trapped states in two-dimensional (2D) and three-dimensional (3D) nonlinear media [1-3], necklace patterns, built as clusters of solitons, pose a challenge due to their structural fragility. Such patterns have been analysed theoretically in diverse models [4-11] and experimentally in Kerr optical media [12]. An inherent feature of necklace clusters is orbital angular momentum, which can be imparted to them [6,9] to induce rotation of the cluster accompanying its radial evolution [8]. Various necklace patterns have been studied in two-component systems [7,9,10]. In most cases clusters are not steady states, as interactions between adjacent solitons induce forces which drive radial oscillations or the eventual self-destruction of the clusters. Indeed, being higher-order states, they are highly prone to instabilities that lead to their splitting into a set of fragments, often through fusion of the initial constituents. Therefore, exploration of physical mechanisms that can support and eventually stabilize such states, which occupy the intermediate niche between multipoles and vortex solitons, is of importance for different fields, including nonlinear optics [4-12] and atomic physics [13,14].

It has been theoretically shown that the decay of clusters is slowed down by competing nonlinearities, such as those represented by a combination of focusing quadratic and defocusing cubic terms, but such conditions are yet to be realized experimentally [10]. Necklaces carrying orbital angular momentum were also explored in models of nonlinear dissipative media based on complex Ginzburg-Landau equations. In that case, the clusters, instead of expansion, may undergo fusion into stable vortex rings [15]. Another possibility for the stabilization of clusters is provided by external potentials, including spatially periodic ones [16]. Such potentials determine the symmetry of the cluster and restrict its motion, by arresting, in particular, radial oscillations. Effective stabilization of necklace patterns was predicted and experimentally explored in nonlocal optical media [17], and also predicted in Bose-Einstein condensates (BECs) with long-range interactions [18]. However, with this exception, all stabilization schemes are awaiting experimental observation, in part due to the difficulty to identify systems, where complex multidimensional states may exist as robust objects.

Recently, a landmark advance in the field has been achieved with the advent of *quantum droplets* (QDs), which were predicted [19] as stable soliton-like states in two-component BECs, with the inter-component attraction made slightly stronger than the intra-component repulsion. A key effect in the system is the Lee-Huang-Yang (LHY) correction to the mean-field approximation, induced by quantum fluctuations [20]. This is a remarkable example of the creation of stable nonlinear objects by competing nonlinearities: cubic mean-field attraction and LHY-induced quartic repulsion. The prediction was followed by the experimental creation of quasi-2D droplets, strongly compressed in one direction [21,22], and 3D isotropic ones [23]. The experiments made use of the Feshbach resonance (FR) to tune interactions between atoms in two hyperfine states [24], so as to make the inter- and intra-species interactions nearly equal in magnitude but opposite in sign. The remaining small imbalance in favor of the inter-species attraction may be used to adjust the relative strength of the LHY corrections and normal cubic terms [21-23]. QDs maintained by the interplay of the LHY terms and long-range attraction in dipolar BEC may be stable as well [25-27]. Formation of droplet crystals in dipolar condensates supported by a harmonic-oscillator trap was recently predicted too [28].

The experimentally relevant [21,22] reduction of the dimension from 3D to 2D changes the form of the LHY terms, replacing the above-mentioned cubic-quartic combination by the cubic terms multiplied by a logarithmic factor [29,30], which implies switching from self-attraction to repulsion with the increase of the density. Such nonlinearity supports stable 2D solitons, including ones with embedded vorticity, $m$, which may be stable up to $m = 5$ [30], while 3D vortex solitons maintained by the cubic-quartic nonlinearity may be stable for $m = 1, 2$ [31]. All vortex solitons supported by the interplay of the long-range dipole interaction and LHY terms are unstable [32].

The goal of this Letter is to show that the 2D geometry allows formation of necklace clusters composed of fundamental droplets with properly selected phase differences, without confining potentials. This prediction opens the way to experimental creation of elusive complex self-sustained states in BEC, which, *inter alia*, exhibit stable oscillations and robust spiralling in free space.

We address the system of coupled Gross-Pitaevskii equations for wave functions $\psi_{1,2}(x,y,t)$ of two components of a binary BEC that builds QDs in two dimensions [29,30]:

$$i\frac{\partial \psi_{1,2}}{\partial t} = -\frac{1}{2}\left(\frac{\partial^2}{\partial x^2}+\frac{\partial^2}{\partial y^2}\right)\psi_{1,2} + (|\psi_{1,2}|^2 - |\psi_{2,1}|^2)\psi_{1,2} + \alpha(|\psi_1|^2+|\psi_2|^2)\psi_{1,2}\ln(|\psi_1|^2+|\psi_2|^2). \quad (1)$$

Here the strength of the cubic difference nonlinearity is scaled to be 1, while $\alpha$ characterizes the strength of the LHY terms. First, we consider the case of constant $\alpha$. Because the FR, controlled by a time-dependent external field, makes it possible to vary the nonlinearity in time [33,34], below we also check the robustness of the necklaces against such variations. Equations (1) conserve norm $U$, energy $E$, and $z$ component $\mathcal{L}$ of the angular momentum:

$$U = \iint (|\psi_1|^2+|\psi_2|^2)dxdy \equiv U_1+U_2,$$
$$E = (1/2)\iint \{|\nabla\psi_1|^2+|\nabla\psi_2|^2+|\psi_1|^4+|\psi_2|^4-2|\psi_1|^2|\psi_2|^2 + \alpha(|\psi_1|^2+|\psi_2|^2)^2[\ln(|\psi_1|^2+|\psi_2|^2)-1/2]\}dxdy, \quad (2)$$
$$\mathcal{L} = (\mathbf{e}_z/2i)\iint [\boldsymbol{\rho}\times(\psi_1^*\nabla\psi_1-\psi_1\nabla\psi_1^*+\psi_2^*\nabla\psi_2-\psi_2\nabla\psi_2^*)]dxdy,$$

where $\boldsymbol{\rho}=(x,y,z)$. We aim to construct ring-shaped clusters as sets of identical QDs. First we elucidate properties of the "building blocks", i.e., fundamental and vortex QDs, generated by Eq. (1) as $\psi_{1,2} = u_{1,2}(r)e^{im_{1,2}\varphi-i\mu_{1,2}t}$, where polar coordinates $(r,\varphi)$ are defined around the centre of a given QD. We look for symmetric stationary states, with real $u_1 = u_2 \equiv u$, vorticities $m_{1,2} \equiv m$, and real chemical potentials $\mu_{1,2} \equiv \mu$. The profiles $u(r)$ of such states, found by means of the Newton method, are not affected by the difference cubic terms in Eq. (1), which allows to set $\alpha = 1$. However, when performing the stability analysis of perturbed states we keep the difference terms in Eq. (1), in order to test the stability of the stationary states against symmetry-breaking perturbations.

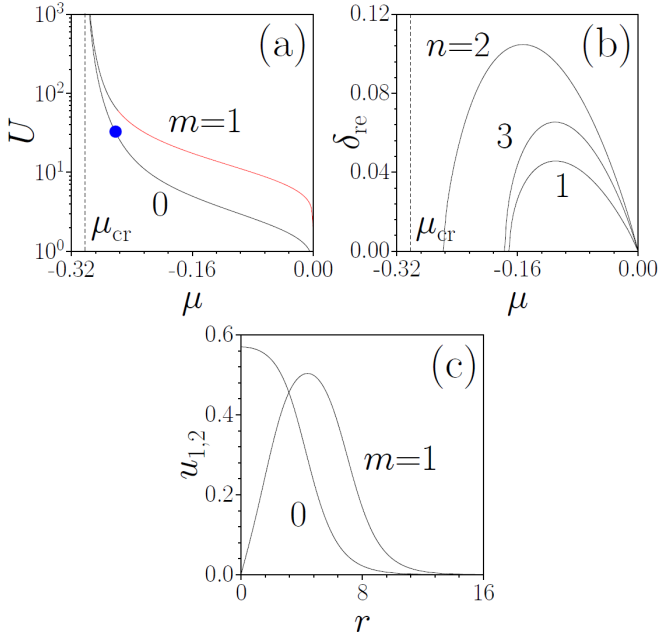

Fig. 1. (a) The norm of the fundamental and vortex QDs vs. the chemical potential. Red and black segments represent stable and unstable states, respectively. The vertical dashed line shows the critical value of $\mu$ below which QDs cease to exist. The blue dot indicates the QD at $\mu = -0.26$, which is used below to construct clusters. (b) The growth rate of perturbations with azimuthal indices, $n$, around the QD with $m=1$, vs. $\mu$. (c) Stable QDs with $m=0$ and $m=1$ at $\mu = -0.26$.

The logarithmic factor in the nonlinear terms lends QDs flat-top profiles, filled by a nearly constant density $\rho \equiv u_{1,2}^2 : \rho_{\text{flat}} = e^{-1/2}$, when $\mu$ approaches the asymptotic value $\mu_{\text{cr}} = \rho_{\text{flat}}\ln\rho_{\text{flat}}$ [see the dashed line in Fig. 1(a)] [30]. Accordingly, the QDs exist for $\mu \in [\mu_{\text{cr}}, 0]$. Their norms are monotonously decreasing functions of $\mu$, vanishing at $\mu \to 0$, see Fig. 1(a). Naturally, the fundamental QD with $m=0$ features the minimal energy $E$ among all states with a fixed norm.

The linear stability analysis of the QDs was performed by substitution in Eq. (1) of perturbed states $\psi_{1,2} = (u_{1,2} + a_{1,2}e^{\delta t+in\varphi} + b_{1,2}^*e^{\delta^* t-in\varphi})e^{-i\mu t+im\varphi}$, where $n$ is the azimuthal perturbation index and $\delta = \delta_{\text{re}} + i\delta_{\text{im}}$ is the perturbation growth rate, followed by the linearization and solution of the resulting eigenvalue problem. For the QDs with $m=0$ and 1 the results are displayed in Fig. 1(a,b). Values of $\mu$ that are close to the stability boundary of $m=1$ QDs turn out to be optimal for constructing clusters made of fundamental QD states – in particular, this is the case of the QD corresponding to the blue dot in Fig. 1(a). The respective necklace clusters are metastable, showing minimal radiation losses at an initial stage of the evolution. The radiation is pronounced if one attempts to build a cluster of fundamental QDs taken too close to $\mu = \mu_{\text{cr}}$, i.e., in the flat-top regime. On the other hand, the clusters built of QDs with too small values of $|\mu|$ are definitely unstable. Radial profiles of QDs for an appropriate $\mu$ are shown in Fig. 1(c), their profiles being far from flat-top ones, with relatively low norms.

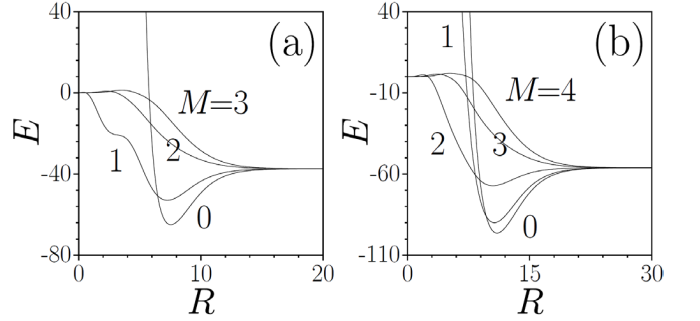

Fig. 2. Energy of the necklace clusters composed of fundamental QDs with $\mu = -0.26$, vs. cluster's radius $R$ for $N=6$ (a) and $N=9$ (b), for overall vorticities $M \leq N/2$ of the cluster.

We construct necklace clusters of $N$ identical QDs placed equidistantly on a ring of radius $R$:

$$\psi_{1,2}|_{t=0} = \sum_{k=1,N} u_{1,2}(|\mathbf{r}-\mathbf{r}_k|)e^{2\pi iMk/N}, \quad (3)$$

where $\mathbf{r}_k = \{R\cos(2\pi k/N), R\sin(2\pi k/N)\}$ is the position of the centre of the $k$-th droplet, and $2\pi M/N$ is the phase difference between adjacent ones. While $M$ may be considered as the cluster's vorticity, the phase varies along the necklaces as a step-like function, making them different from ordinary vortex states. Actually, $M$ is a key factor controlling the evolution of the cluster, as it determines interaction forces between adjacent QDs via phase difference $\theta = 2\pi M/N$ between them. Namely, for $M \leq N/4$ and $M \geq 3N/4$ the interaction between adjacent QDs is attractive, corresponding to $\theta \leq \pi/2$ or $\theta \geq 3\pi/2$, hence the net force, pointing at the centre, leads to an initial shrinkage of the cluster in the radial direction. For $N/4 < M < 3N/4$, the phase difference between adjacent QDs takes values $\pi/2 < \theta < 3\pi/2$, giving rise to a repulsive net force in the radial direction, that causes expansion of the cluster. Moreover, at $M \neq 0$ and $M \neq N/2$ the cluster carries nonzero angular momentum $\mathcal{L}$, hence its radial contraction/expansion is accompanied by rotation.

The input state (3) may be prepared by slicing the BEC in the trap by several narrow repulsive barriers, similar to how it was done for two droplets [35]. Phase-imprinting techniques, implemented by means of a far-detuned broad vortical beam [36], may be used to imprint the required phases onto the droplets forming the cluster.

One can estimate the expected evolution regimes of the clusters from the dependence of energy $E$ on the initial cluster's radius, $R$, which is shown in Fig. 2 for different numbers $N$ of QDs in the cluster. These dependencies reveal that at $M \leq N/4$ an energy minimum exists at a specific initial radius $R_{\min}$. The state realizing the energy minimum is a favorable one, hence it should exhibit minimal shape changes in the course of evolution. Clusters with $R > R_{\min}$ perform nearly periodic oscillations [Figs. 3(a,b)], clearly seen in the time dependence of the cluster radius, $w = U^{-1} \iint r(|\psi_1|^2 + |\psi_2|^2) dx dy$, in Fig. 3(b). For large $R$, corresponding to the region, where $E$ is nearly constant, the oscillation period is very large too. For $M = 0$, conspicuous emission of radiation occurs at the point of the maximal contraction of the cluster, gradually transforming it into a nearly flat-top and practically non-radiating pattern, that inherits the initial azimuthal structure [Fig. 3(a)]. At $\mathcal{L} \neq 0$, the radial oscillations are accompanied by rotation of the cluster. Clusters with $R < R_{\min}$ initially expand and then also show oscillations (see an example in Fig. 5(a) at $t > 200$).

The central result of this Letter is the robustness of the quasi-stationary QD clusters in comparison with media with the usual cubic nonlinearity, where clusters are strongly unstable [12]. To elucidate their robustness, we added different random perturbations (up to 5% in amplitude) to the initial profiles of the two components. After performing several cycles of radial oscillations, the perturbed clusters with $R$ essentially different from $R_{\min}$ split into several wide droplets, a phenomenon that is usually accompanied by a strong growth of the width, as shown by the red curve in the top row of Fig. 3(b). On the contrary, clusters with initial radii $R \approx R_{\min}$ are robust against the perturbations, surviving up to $t > 2000$, as seen in the top panel of Fig. 3(c). Typical metastable clusters are shown in Fig. 4, with their radius increasing with $N$. We have verified that the minima in the $E(R)$ dependencies in Fig. 2 provide an accurate prediction of $R_{\min}$, only slightly underestimating (by ~5%-7%) the actual value of the radius at which the clusters exhibit no appreciable oscillations.

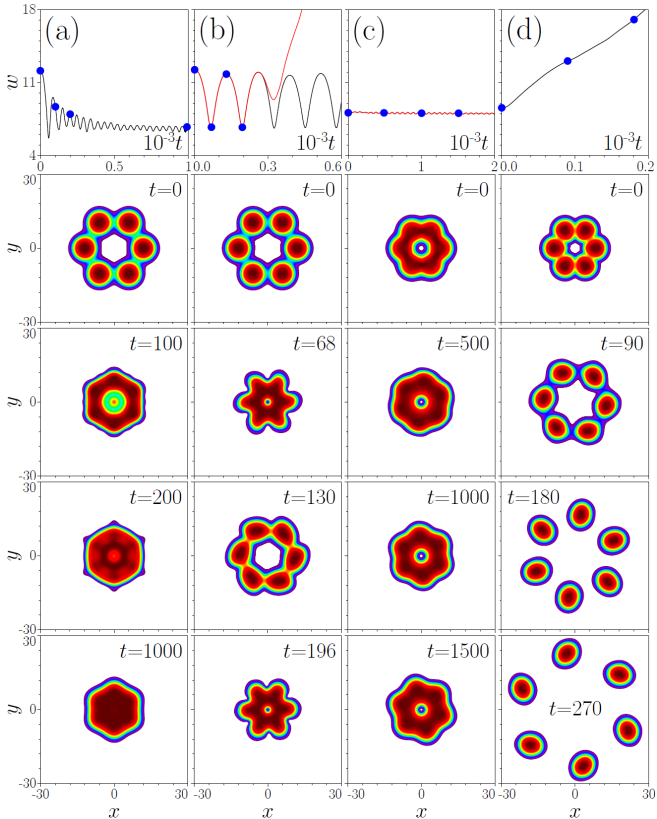

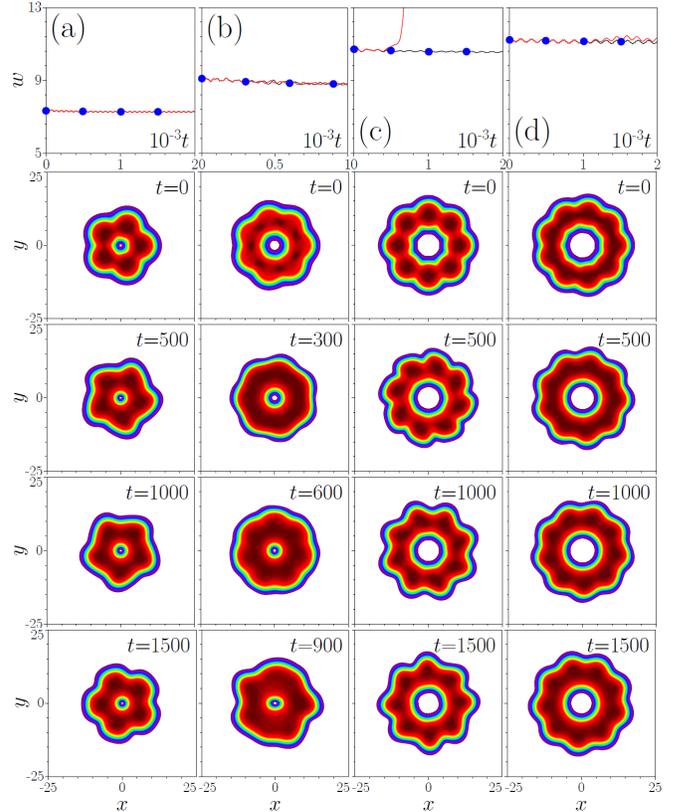

Fig. 3. Different regimes of the evolution of clusters with $N = 6$. (a) Gradual fusion into a broad fundamental state, at $M = 0$, $R = 12$; (b) periodic oscillations, at $M = 1$, $R = 12$; (c) evolution with a nearly constant radius, at $M = 1$, $R = 7.9$; (d) expansion, at $M = 2$, $R = 7.9$. The top row shows the evolution of the cluster's radius. Profiles $|\psi_1|$ are shown at different moments of time corresponding to blue dots in the $w(t)$ dependencies. Red (black) curves in the first row correspond to explicitly perturbed (unperturbed) clusters, see the text. In panels (a), (c), and (d), the evolution of perturbed and unperturbed clusters is indistinguishable.

The most interesting situation takes place at $R \approx R_{\min}$. In this regime, there emerge quasi-stationary states which perform persistent rotation with minimal radial oscillations [Fig. 3(c)]. The range of values of $M$ where the formation of such states is possible expands with the increase of the number of droplets $N$ in the cluster [cf. Fig. 2(b) and 2(a)]. Finally, for clusters with $N/4 < M < 3N/4$ the global energy minimum is achieved only at $R \to \infty$, hence they expand due to the repulsion between adjacent QDs [Fig. 3(d)].

Fig. 4. Metastable evolution of droplet clusters with nearly constant radii for (a) $N = 5$, $M = 1$, $R = 7$, (b) $N = 7$, $M = 1$, $R = 8.85$, (c) $N = 8$, $M = 2$, $R = 10.4$, and (d) $N = 9$, $M = 2$, $R = 11$. The meaning of the top row is the same as in Fig. 3.

The numerically found optimal values of the radius were used in all simulations of the evolution of the clusters displayed in Fig. 4. Most robust are ones with $N = 5$ [Fig. 4(a)] and $N = 6$ [Fig. 3(c)]. For other states, the evolution time over which the perturbed cluster survives usually decreases with $N$. All clusters in Fig. 4 exhibit rotation, which is slower than small radial pulsations in the $w(t)$ dependencies. A representative rotation period of the $N = 5$ cluster in Fig. 4(a) is $T \approx 245$. The present results persist in the presence of weak harmonic confinement in the $(x, y)$ plane, that may even enrich families of metastable clusters [37].

In addition to the stability of the clusters against initial perturbations, we have tested their robustness with respect to variation of parameters. In particular, this may be abrupt or smooth change of the nonlinearity strength, imposed by FR. To this end, we took a robust cluster with $N = 5$ and checked how a change of $\alpha$ in Eq. (1) at

$t \geq 200$ affects the evolution. While for smooth variation of $\alpha$ the cluster's radius adapts to it, an abrupt jump from $\alpha = 1$ to a smaller value entails noticeable radial oscillations, with the minimal radius being determined by the initial one [Fig. 5(a)]. If $\alpha$ increases, the cluster shrinks instead, with the initial radius determining the maximal radius of subsequent oscillations [Fig. 5(b)]. The dependence of the maximal and minimal radii of the cluster on the final value of $\alpha$ is displayed in Fig. 6(a). Note that the so perturbed clusters survive even if their radii undergo more than a double increase in the course of oscillations. There is a lower bound on the possible variation of $\alpha$, since at $\alpha < 0.2$ the cluster expands indefinitely instead of performing radial oscillations, but no upper bound exists. The oscillation period monotonically decreases with $\alpha$ [Fig. 6(b)].

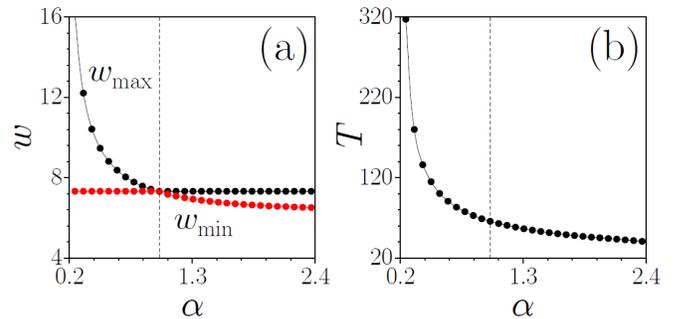

Fig. 6. Maximal and minimal radii (widths) of the necklace cluster (a) and its oscillation period (b) versus final nonlinearity strength $\alpha$, see Eq. (1). At $t = 0$ the cluster was created with $N = 5$, $M = 1$, $R = 7$. Dashed lines correspond to $\alpha = 1$, i.e. to constant nonlinearity.

We conclude that remarkably robust, albeit, rigorously speaking, metastable ring-shaped clusters built of QDs may be formed under suitable conditions in binary BEC in 2D geometries, where the LHY corrections to the mean-field dynamics provide the stabilization. The evolution of the clusters is determined by their initial radius and vorticity. The clusters selected by the energy-minimum principle withstand strong perturbations, such as the variation of the nonlinearity strength.

L.T. and Y.V.K. acknowledge support from the Severo Ochoa program (SEV-2015-0522) of the Government of Spain, Fundacio Cellex, Fundació Mir-Puig, Generalitat de Catalunya, and CERCA. Y.V.K. acknowledges partial support of this work by program 1.4 of Presidium of the Russian Academy of Sciences, "Topical problems of low temperature physics." The work of B.A.M. is supported, in part, by the Israel Science Foundation through grant No. 1287/17.

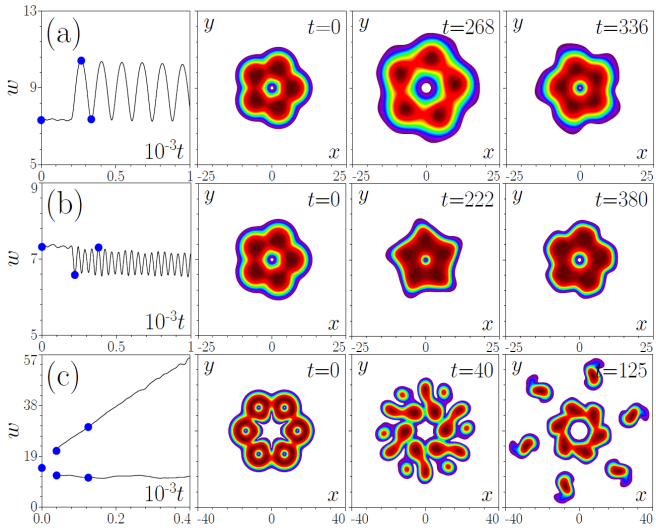

Fig. 5. Oscillations of the radius of the quasi-stationary cluster with $N = 5$, $M = 1$, $R = 7$, induced by an abrupt jump of the nonlinearity coefficient in Eq. (1) from $\alpha = 1$ to $\alpha = 0.4$ (a) or $\alpha = 2$ (b) at $t = 200$. (c) The evolution of the cluster with $R = 14$ and overall vorticity $M = 1$, composed of 6 QDs with $\mu = -0.26$ and intrinsic vorticity $m = 1$. Widths of the quasi-stationary internal and expanding external clusters are shown starting from a time moment at which they are split.

Necklace clusters can be built not only of fundamental QDs with $m = 0$, but also, as *supervortices* [16], of compact vortical QDs with $m = 1$. An example of such a cluster and its evolution are shown in Fig. 5(c). The phase counterflow at junctions between adjacent vortices initiates their considerable reshaping, through a mechanism akin to the Kelvin-Helmholtz instability [38]. In the case displayed in Fig. 5(c), this results in the formation of a nearly stationary rotating inner cluster and expanding external one, both composed of fundamental droplets. Radii of the clusters are shown as functions of time in the left panel of Fig. 5(c), starting from the moment when they clearly separate. This outcome implies approximate cancellation of the cluster's orbital momentum and the net internal momenta of the constituents (computation shows that $75\%$ of the total momentum is cancelled in the input). Reversing the overall vorticity to $M = -1$ does not allow the cancellation and leads to a different outcome, breaking the input into fragments (not shown here). Attempts to build a necklace of droplets with alternating values of the inner vorticity, $m = \pm 1$, result in quick destruction of the patterns.